\documentstyle[multicol,aps,prl,eqsecnum,epsf]{revtex}

\newcommand{\ave}[1]{\left\langle #1 \right\rangle}

\newcommand{\eqn}[2]{\begin{equation}#2\label{#1}\end{equation}}

\newcommand{\eqna}[2]{\begin{eqnarray}#2\label{#1}\end{eqnarray}}   %%

\begin{document}
\title{A new Eulerian-Lagrangian length-scale in turbulent flows}
\author{M. A. I. Khan  \\
\small{D.A.M.T.P., University of Cambridge, Silver Street, 
Cambridge CB3 9EW, U.K.}\\
J.C. Vassilicos\\
\small{Department of Aeronautics, Imperial College, London SW7 2BY, U.K.}}
\maketitle
\begin{abstract}

We introduce a time-dependent Eulerian-Lagrangian length-scale and
an inverse locality hypothesis which explain scalings of second
order one-particle Lagrangian structure functions observed in
Kinematic Simulations (KS) of homogeneous isotropic turbulence. Our KS
results are consistent with the physical picture that particle
trajectories are more/less autocorrelated if they are smoother/rougher
as a result of encountering less/more straining stagnation points,
thus leading to enhanced/reduced turbulent diffusion.

\end{abstract}
%\pacs{}
%\keywords{turbulent diffusion}
\begin{multicols}{2}

\section{Introduction}
Turbulent diffusion is a phenomenon of central importance in oceanic,
atmospheric, engineering and astrophysical flows. It is known, since
the seminal work of Taylor \cite{taylor21}, that the turbulent diffusivity (of
contaminants, pollutants or other substances  advected by the
turbulent flow) is proportional to the time-integral of the Lagrangian
velocity autocorrelation function $R_{L}(\tau)$ of the turbulence. The
Lagrangian study of fluid element trajectories is therefore central to
understanding and calculating turbulent diffusion.

A consequence of Taylor's (1921) \cite{taylor21} theory is that the turbulent
diffusivity is larger/smaller depending on whether Lagrangian
velocities are correlated over longer/shorter times along
trajectories. Such long/short correlation times reflect large/small
Lagrangian velocity variations, and Lagrangian velocity variations
over a time interval $\tau$ are characterized by the second order
Lagrangian structure function which is proportional to $1 -
R_{L}(\tau)$. In statistically stationary and isotropic turbulence it
is sufficient to consider only one component of the Lagrangian
velocity, say $v(t)$; its second order structure function is
$\ave{\delta v^{2}(\tau)} \equiv \ave{(v(t) - v(t+\tau))^{2}}$
corresponding to the Lagrangian velocity power spectrum
$E_{L}(\omega)$ in the frequency ($\omega$) domain. As noted by Inoue
\cite{Inoue51} (see also Tennekes \& Lumley \cite{tennekes72}), Kolmogorov similarity
arguments imply that, in the inertial range of times and frequencies,
$\ave{\delta v^{2}(\tau)} \sim \epsilon \tau$ and $E_{L}(\omega) \sim
\epsilon \omega^{-2}$ where $\epsilon$ is the kinetic energy
dissipation rate per unit mass. Support for these scalings has
recently been obtained in highly turbulent oceanic environments by
Lien {\it et al.} \cite{lien98}, in the laboratory by Mordant {\it et al.}
\cite{mordant2002} and in Direct Numerical Simulations (DNS) by Yeung
\cite{yeung2001}. However, Kolmogorov similarity arguments provide very little
understanding of the underlying processes responsible for these
scaling laws. What properties of the spatio-temporal flow structure of
the turbulence determine that $\ave{\delta v^{2}(\tau)}$ is
proportional to $\tau$, and what do these properties imply for
turbulent diffusion?

In this letter we address this question by use of Kinematic Simulation
(KS, see Nicolleau \& Vassilicos \cite{nicolvas2002}, Davila \& Vassilicos \cite{davila2002} and
references therein). KS is a Lagrangian model of turbulent diffusion
based on kinematically simulated turbulent velocity fields which are
incompressible and consistent with up to second order Eulerian
statistics of the turbulence such as energy spectra $E(k)$ in the
wavenumber ($k$) domain. There is no assumption of Markovianity or
delta correlation in time made at any level. Instead, a parameter
$\lambda$ controls the degree of unsteadiness of the turbulence. It
might be worth mentioning that, when the energy spectrum input has the
Kolmogorov $k^{-5/3}$ form, KS is in good agreement with laboratory
experiments for one-particle statistics \cite{osborne2002},
two-particle statistics \cite{nicolvas2002}, three-particle
statistics \cite{kpv2002} and concentration variances \cite{FlohrV2000} (particle is used here to mean fluid element). KS is 
also in good agreement with DNS for two-particle statistics when the KS 
energy spectrum has the same overall shape as that of the DNS 
\cite{malik99}. In this letter, we modify the spatio-temporal 
flow structure of the KS turbulence by changing the input energy spectrum 
$E(k)$ and the input unsteadiness parameter $\lambda$ and study how 
$\ave{\delta v^{2}(\tau)}$ changes as a result.  

\section{Kinematic simulation}
In our KS we use 3D turbulent-like velocity fields of the form
\vspace*{-7mm}
\eqna{KS}{ {\bf u} = \sum_{n=1}^{N_k} {\bf A}_n\wedge\hat{\bf k}_n
   \cos({\bf k}_n\cdot{\bf x} +  \omega_n t) & + \nonumber \\ {\bf
   B}_n\wedge\hat{\bf k}_n \sin({\bf k}_n\cdot{\bf x} +  \omega_n t) &
   }
where $N_k$ (typically of order 100) is the number of modes, $\hat{\bf
k}_n$ is a random unit vector (${\bf k}_n = k_n \hat{\bf k}_n$), and
the directions and orientations of ${\bf A}_n$ and ${\bf B}_n$ are
chosen randomly under the constraint that they be normal to $\hat{\bf
k}_n$ and uncorrelated with the directions and orientations of all
other wave modes. Note that the velocity field (\ref{KS}) is
incompressible by construction, and also statistically stationary,
homogeneous and isotropic as shown by Fung {\it et al.} \cite{FHMP92} and
Fung \& Vassilicos \cite{fungvas98}.  The amplitudes $A_n$ and $ B_n$ of the
vectors ${\bf A}_n$ and ${\bf B}_n$ are determined by $ A_n^2 = B_n^2
= \frac{2}{3} E(k_n) \Delta k_n $ where $ E(k) $ is the energy
spectrum prescribed to be of the form 
\eqn{energ}{ E(k) =
\frac{v'^{2}(p-1) }{ 2(L/2\pi)^{p-1} } \, k^{-p} } 
in the range $
2\pi/L = k_1 \le k \le k_{N_k} = 2\pi/\eta $ assumed large
(i.e. $L/\eta \gg 1$), and $ E(k)=0 $ otherwise; $ v' $ is the rms of
one velocity component of the KS turbulent-like flow; $ \Delta k_n =
(k_{n+1} - k_{n-1})/2 $ for $ 2 \le n \le N_k - 1 $, $ \Delta k_1 =
(k_2 - k_1)/2 $ and $ \Delta k_{N_k} = (k_{N_k} - k_{N_k-1})/2 $. The
distribution of wavenumbers is geometric (see \cite{FlohrV2000}), specifically $ k_n = k_1 a^{n-1} $ with a constant $a$
determined by $ L/\eta = a^{N_k -1} $. The frequencies $\omega_n$ in
(\ref{KS}) are proportional to the eddy-turnover frequency of mode
$n$, and the dimensionless constant of proportionality is $\lambda$,
i.e.  $\omega_n = \lambda\sqrt{k_n^3 E(k_n)}$.
%----------------figure----------------------
\narrowtext \begin{figure} \epsfxsize=8truecm 
\epsfbox{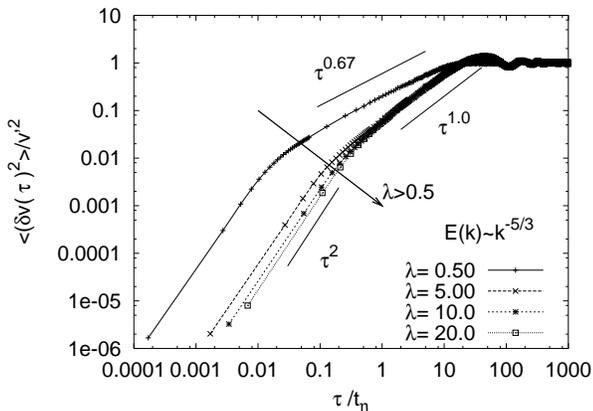}
\caption{The time evolution of the velocity increment $\ave{\delta 
v^{2}(\tau)}$ in a log-log plot.}
\label{f1}
\end{figure}
%-----------------figure--------------------
Particle trajectories are numerically integrated in (\ref{KS}) and the
velocities of the particles are recorded along their flight. By averaging
over $5\times 10^{3}$ trajectories in $5\times 10^{3}$ realizations of the
velocity field we obtain $\ave{\delta v^{2}(\tau)}$ for different values
of $p$ and $\lambda$ (see Fig. \ref{f1} which has been obtained for 
$L/\eta =
1690$, but note that we corroborated these results for much higher values
of $L/\eta$ too). We try values of $p$ larger than 1 to ensure that the
total kinetic energy of the turbulence is finite in the large $L/\eta$
limit, and values of $p$ smaller than 2 for the power spectrum $k^{2}E(k)$
of the vorticity and strain rate fields to be an increasing function of
$k$ (see \cite{davila2002}).

\section{Results}
A first observation is that when $E(k)$ has the Kolmogorov similarity
dependence on wavenumber (i.e. $p=5/3$), $\ave{\delta v^{2}(\tau)}$
exhibits the Kolmogorov similarity scaling $\ave{\delta v^{2}(\tau)}
\sim \tau$ over an intermediate range of time-scales only for large
enough values of $\lambda$ and not otherwise. This intermediate range
of scales turns out to be $t_{\eta}\ll \tau \ll T_{L}$ where
$t_{\eta}=2\pi/\omega_{N_{k}}$. For values of $\lambda$ smaller than 1
and close to 0 as well as equal to 0, we find $\ave{\delta
v^{2}(\tau)} \sim \tau^{2/3}$ over the intermediate range
$\tau_{\eta}\ll \tau \ll T_{L}$ where $\tau_{\eta} \equiv
2\pi/\sqrt{(2\pi/\eta)E(2\pi/\eta)}$ and $T_{L}\equiv
\int_{0}^{\infty} R_{L}(\tau) d \tau$ ($t_{\eta}=\tau_{\eta}/\lambda$). 
Fung {\it et al.} \cite{FHMP92} have already observed
this 2/3 scaling in frozen ($\lambda =0$) KS turbulence and added to
(\ref{KS}) a sweeping of the small-scales (\ref{KS}) by larger
small-scales which restored the Kolmogorov similarity scaling
$\ave{\delta v^{2}(\tau)} \sim \tau$. Hence, large values of the
unsteadiness parameter $\lambda$ reproduce this sweeping effect on the
scaling of $\ave{\delta v^{2}(\tau)}$. Kolmogorov similarity is
conditional on large values of the {\it dimensionless} parameter
$\lambda$, perhaps a surprising result.

%---------------figure--------------------
\narrowtext \begin{figure} \epsfxsize=8truecm
\epsfbox{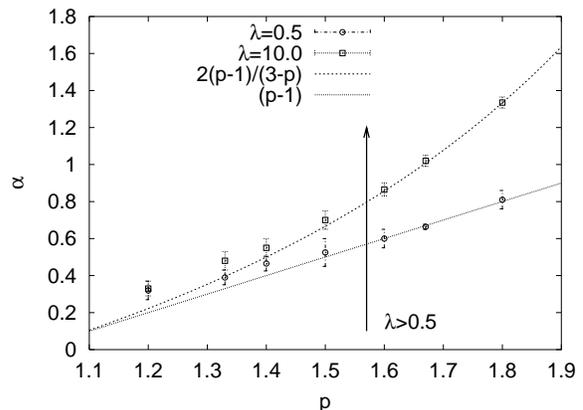}
\caption{Plot of the change in the power law of the Lagrangian velocity
increment ($\langle\delta v(\tau)^{2}\rangle\sim \tau^{\alpha}$ as a 
function of $p$) in KS for different $\lambda$'s. The exponent $\alpha$ and
its error bars are calculated from a least square fit over the range of 
scales where
$\langle\delta v(\tau)^{2}\rangle\sim \tau^{\alpha}$ is observed. 
This power
law is harder to detect accurately for values of $p$ closer to 1 because a 
larger proportion of the energy is then distributed to higher wavenumbers 
and both $\tau_{\eta}$ and $t_{\eta}$ increase indefinitely as $p \to 1$ thus
requiring an ever larger ratio $L/\eta$ and therefore value of $T_{L}$. The
fact that more energy is distributed at the higher wavenumbers when
$p$ decreases
means that trajectories are more irregular, a fact reflected in the smaller
values of $\alpha$. With our extremely good time resolutions and 
ratios $L/\eta$ up to $10^4$ we still cannot accurately measure values 
of $\alpha$ below about 0.4.}   
\label{f2}
\end{figure}
%---------------figure-------------------
Running simulations for a variety of values of $p$ between 1 and 2 we
find 
\eqn{alpha}{ \ave{\delta v^{2}(\tau)} \sim \tau^{\alpha} } 
in the 
intermediate ranges mentioned above and $\alpha = \alpha (p, \lambda)$
(see figures \ref{f1} and \ref{f2}). For
values of $\lambda$ smaller than 1 including $\lambda = 0$,
\eqn{lambdasmall}{ \alpha = p-1 ,} 
but for values of $\lambda$
much larger than 1, $\alpha$ is a monotonically increasing
nonlinear function of $p$, in fact the same function of $p$ for all large
enough $\lambda$, and is such that $\alpha > p-1$ and $\alpha =1$
at $p=5/3$. What is this function of $p$? 

To answer this question we proceed by analogy with Richardson's
locality hypothesis for two-particle diffusion in statistically
homogeneous turbulence. According to this hypothesis, the rate of
change of the mean square separation between two particles, i.e.
${d\over dt} \ave{ \Delta^{2}}$, is a function only of $\ave{
\Delta^{2}}$ and of the variance of Eulerian velocity variations
across that length-scale (see \cite{fungvas98}). That is
\eqn{locality} { {d\over dt} \ave{ \Delta^{2}} = f \left( \ave{
\Delta^{2}}, \ave{\delta u^{2} (\sqrt{\ave{ \Delta^{2}}})}\right)}
where $\ave{\delta u^{2} (r)}$ is the second order structure function
of one component of the Eulerian velocity field. It is well known that
$\ave{\delta u^{2} (r)} \sim r^{p-1}$ for small enough values of $r$
is equivalent to $E(k)\sim k^{-p}$ for large enough values of $k$.
The analogy we are trying here, which we call the inverse locality
hypothesis, is to say that the rate of change of $\ave{\delta
v^{2}(\tau)}$ with respect to $\tau$ is a function only of
$\ave{\delta v^{2}(\tau)}$ and of the length-scale across which the
variance of Eulerian velocity variations is equal to $\ave{\delta
v^{2}(\tau)}$. That is 
\eqn{invlocality}{ {d\over d\tau}\ave{\delta
v^{2}(\tau)} = f\left (\ave{\delta v^{2}(\tau)}, l_{e}(\tau)\right) }
where $l_{e}(\tau)$ is an Eulerian-Lagrangian length-scale defined as
follows:  
\eqn{le}{ \ave{\delta u^{2}(l_{e}(\tau))} = \ave{\delta
v^{2}(\tau)}.} 
We call $l_{e}(\tau)$ the equivalent length-scale.
Dimensional analysis applied to (\ref{invlocality}) and use of
(\ref{le}) and $\ave{\delta u^{2} (r)} \sim r^{p-1}$ leads to
(\ref{alpha}) with  
\eqn{lambdalarge}{ \alpha = {2(p-1)\over 3-p}.}
This dependence of $\alpha$ on $p$ fits surprisingly well the results
obtained by KS for large values of $\lambda$ (see Fig. \ref{f2}). These
results reveal, therefore, the existence of an Eulerian-Lagrangian
length-scale $l_{e}(\tau)$ which controls the second order Lagrangian
structure function and Lagrangian frequency spectrum provided that the
sweeping effects are taken into account by high values of $\lambda$.
Our results (figures \ref{f1} and \ref{f2} and equations (\ref{lambdasmall}) and (\ref{lambdalarge})) also reveal
the  impact that the two-point spatial flow structure (as reflected in
the  scaling of the energy spectrum) has on one-particle Lagrangian
statistics whether with or without sweeping.

\begin{figure} \epsfxsize=8truecm
\epsfbox{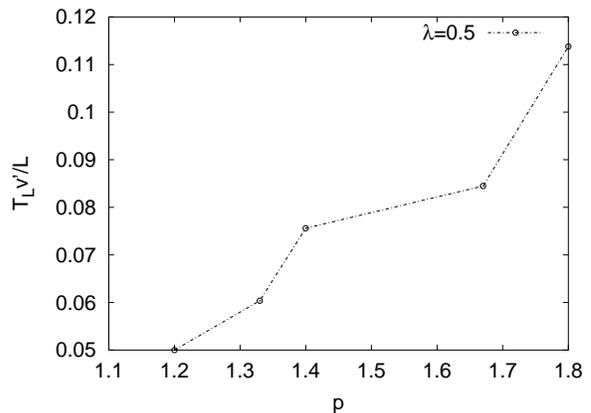}
\caption{Plot of Lagrangian correlation time $T_{L}$ as a function of
$p$ for $\lambda=0.5$ in KS with $L/\eta=100$. Similar results are found 
for other values of $\lambda$ and $L/\eta$.}
\label{T_L_lm5_10}
\end{figure}

Changes in $\ave{\delta v^{2}(\tau)}$ imply changes in $R_{L}(\tau)$
and therefore in turbulent diffusion. Taylor's (1921) formula for the
mean square  one-particle displacement $\ave{x^{2} (t)}$  is
\eqn{}{ {d\over dt}\ave{x^{2} (t)} = v'^{2}\int_{0}^{t}
R_{L}(\tau)  d \tau} 
which implies a turbulent diffusivity equal to
$v'^{2}T_{L}$. Of course this relation remains valid in our KS for any
value of $p$ but the Lagrangian correlation time scale $T_{L}$ is
found to be an increasing function of $p$ (see Fig. \ref{T_L_lm5_10}
), thus implying a  turbulent  diffusivity that is also an increasing
function of $p$. This conclusion can be explained qualitatively by the
relation $p+2D_{s}/3 =3$ between $p$ and the fractal dimension $D_{s}$
of the set of straining stagnation points that has recently been
obtained by Davila \& Vassilicos \cite{davila2002} for KS
turbulence. As $p$ decreases, $D_{s}$ increases implying an increase
in the number density of straining stagnation points and thereby an
increased probability for particle trajectories to come in the
vicinity of such points and experience sudden changes of tack. As a
consequence, the Lagrangian correlation time scale is smaller/larger
for smaller/larger values of $p$. In terms of $D_s$, $\alpha = 2
{3-D_{s}\over D_{s}}$ when $\lambda$ is large and $\alpha = {2\over
3}(3-D_{s})$ when $\lambda$ is small; $\alpha$ is therefore a
decreasing function of $D_{s}$ thus reflecting the fact that a larger
number of changes of tack produces more irregular particle
trajectories.

%-----------figure------------------
\narrowtext 
%-----------figure------------------

\section{Acknowledgements}
MAIK and JCV acknowledge support from EPSRC grant GR/K50320, from EC
TMR Research network on intermittency in turbulent systems and from
the Hong Kong Research Grant Council, China (Project
No. HKUST6121/00P). MAIK also wishes to thank the Cambridge
Commonwealth Trust, Wolfson College, Cambridge and DAMTP for financial
support while this work was being completed. JCV is grateful to the
Royal Society of London for its unflinching support over the past
eight years.

\end{multicols}

\begin{references}
\bibitem{taylor21}
G. I .Taylor, Diffusion by continuous movements,
\emph{Proc. Lond. Math. Soc.}, {\bf 2}(20), 196(1921).
\bibitem{Inoue51}
E. Inoue, On turbulent diffusion in the
atmosphere. J. Met. Soc. Japan {\bf 29}, 246(1951).   
\bibitem{tennekes72}
H. Tennekes and J. L. Lumley, A first course in
Turbulence. Cambridge, MA: MIT Press, 1972.
\bibitem{lien98}
R. -C. Lien, E. A. D'Asaro and G. T. Dairiki, Lagrangian spectra of
vertical velocity and vorticity in high Reynolds number oceanic
turbulence. J. Fluid Mech. {\bf 362}, 177(1998).
\bibitem{mordant2002}
N. Mordant, P. Metz, O. Michel and J. F. Pinton, Measurement of
Lagrangian velocity in fully developed turbulence,    
\emph{Phys. Rev. Lett}, {\bf 87}(21), 214501(2002).
\bibitem{yeung2001}
P. K. Yeung, Lagrangian characteristics of turbulence and scalar
transport in direct numerical simulations. J. Fluid Mech. {\bf 427},
241(2001).

\bibitem{nicolvas2002}
F. Nicolleau and J. C. Vassilicos, Turbulent Pair
Diffusion (LANL archive: nlin.CD/0205003).
\bibitem{davila2002}
J. Davila and J. C. Vassilicos, Richardson pair diffusion and the
stagnation point structure of turbulence (LANL archive: physics/0207108).
\bibitem{osborne2002}
D. Osborne, J. C. Vassilicos and J. Haigh (In preparation). 
\bibitem{kpv2002}
M. A. I. Khan, A. Pumir and J. C. Vassilicos, Geometry of multipoint
turbulent dispersion (In preparation).
\bibitem{FlohrV2000}
P. Flohr and J. C. Vassilicos, A scalar subgrid model with flow structure
for large-eddy simulations of scalar variances, \emph{J. Fluid. Mech},
{\bf 407}, 315(2000).
\bibitem{malik99}
N. Malik and J. C. Vassilicos, A Lagrangian model for turbulent
dispersion with turbulent-like flow structure: comparison with direct
numerical simulation for two-particle statistics, \emph{Phys. Fluids.},
{\bf 11}(6), 1572(1999).
\bibitem{FHMP92}
J. C. H. Fung, J. C. R. Hunt, N. Malik and R. J. Perkins, Kinematic
simulation of homogeneous turbulence by unsteady random Fourier modes,   
\emph{J. Fluid. Mech.}, {\bf 236}, 281(1992).
\bibitem{fungvas98}
J. C. H. Fung and J. C. Vassilicos, Two-particle Dispersion in
Turbulent like Flows, \emph{Phy. Rev. E}, {\bf 57}(2), 1677(1998).
\end{references}
\end{document}